\documentclass[aps,twocolumn,superscriptaddress,nofootinbib]{revtex4-2}

\usepackage{graphicx,color}
\usepackage{amsmath,amssymb,bm}
\usepackage{dcolumn}
\usepackage{comment}

\newcommand{\bea}{\begin{eqnarray}}
\newcommand{\eea}{\end{eqnarray}}
\newcommand{\be}{\begin{equation}}
\newcommand{\ee}{\end{equation}}
\newcommand{\dif}{{\rm d}^3}

\newcommand{\np}{{\bf p}}
\newcommand{\nP}{{\bf P}}
\newcommand{\nr}{{\bf r}}

\newcommand{\nh}{{\bf h}}

\newcommand{\nk}{{\bf k}}

\newcommand{\nq}{{\bf q}}

\newcommand{\nj}{{\bf j}}
\newcommand{\nJ}{{\bf J}}
\newcommand{\nR}{{\bf R}}

\newcommand{\nsigma}{\mbox{\boldmath $\sigma$}}
\newcommand{\ntau}{\mbox{\boldmath $\tau$}}

\begin{document}

\title{Short-Range Correlations and Transverse Response
  Enhancement from Meson-Exchange Currents}

\author{P.R. Casale} \email{palomacasale@ugr.es}
\author{J.E. Amaro}\email{amaro@ugr.es} 
\affiliation{Departamento de
  F\'{\i}sica At\'omica, Molecular y Nuclear,
 Universidad de Granada,  E-18071 Granada, Spain.}
\affiliation{Instituto Carlos I
  de F{\'\i}sica Te\'orica y Computacional,
 Universidad de Granada,  E-18071 Granada, Spain.}

\begin{abstract}
We investigate the role of short-range correlations (SRC) in the
transverse nuclear response within the quasielastic peak region,
focusing on the 1p1h channel. The calculation is performed in nuclear
matter by solving the Bethe–Goldstone equation with the realistic
Granada 2013 nucleon–nucleon potential, including both one-body and
two-body meson-exchange currents (MEC) of seagull, pion-in-flight, and
$\Delta$ types. We find that MEC produce a sizable enhancement of the
transverse response in the 1p1h channel when acting on correlated
nucleon pairs with high-momentum components generated by SRC. This is
in contrast to the uncorrelated case, where MEC—dominated by the
$\Delta$ current—can even yield a negative effect.  The results are
consistent with previous findings based on the correlated basis
function approach, supporting the interpretation that SRC play a
central role in the transverse response enhancement from MEC.

\end{abstract}



\maketitle

\section{Introduction}

A long-standing issue in inclusive electron--nucleus scattering is the
excess strength observed in the transverse response function compared
to theoretical predictions based on the impulse
approximation~\cite{Jou96}. Models that accurately reproduce the
longitudinal response---dominated by the charge
operator---systematically underestimate the transverse channel unless
an additional enhancement mechanism is introduced~\cite{Ama94b}. An
empirical parametrization of the required enhancement in $^{12}$C has
been extracted in~\cite{Bod22,Bod24,Bod24b} from global analyses of
inclusive electron scattering data. This transverse excess is most
clearly visible in the central region of the quasielastic peak,
particularly at moderate momentum transfers ($q$ between 300 and
570~MeV/c), where the dominant process is single--nucleon emission. In
contrast, multi--nucleon and pion--production processes contribute
mainly to the high--energy tail of the cross section and cannot
account for the observed strength near the peak. The need for
enhancement in this kinematic region has often been attributed to
meson--exchange currents (MEC), a conclusion supported by \textit{ab
  initio} calculations in nuclei with $A\leq 12$ \cite{Car02,Lov16}. 
These studies
demonstrate that interference between one-- and two--body currents
plays a crucial role in reproducing the experimentally observed
enhancement in the electromagnetic transverse response
function~\cite{And24}, and should not be neglected in
calculations of nuclear responses~\cite{Bar21}. In particular, this
has important implications for neutrino--nucleus scattering and the
modeling of nuclear effects in Monte Carlo event
generators~\cite{Bar21,Lov14,Alv18,Ank22}.

Within single-particle approaches such as the Fermi gas (FG) or the
shell model ---supplemented when necessary with corrections including
random phase approximation (RPA) correlations and final-state
interactions (FSI)--- it is possible to reproduce the gross features of
the quasielastic cross section
\cite{Cen01,Bof96,Mon71,Hor90,Weh93,Sar93}.  
However, these models fail to account for the transverse enhancement
attributed to MEC, and instead tend to exhibit a negative interference
between one- and two-body currents
\cite{Koh81,Alb90,Ama94a,Ama03,Cas23b}.  
This behavior appears to be a general feature, as recently confirmed
in an intercomparison of single-particle models \cite{Cas25}.  
MEC are typically described in terms of one-pion exchange, with the
leading contributions arising from the seagull, pion-in-flight, and
$\Delta$-excitation operators.  
The origin of the negative interference lies primarily in the
$\Delta$ current, which becomes dominant at larger momentum transfers
and interferes destructively with the one-body current, thereby
partially cancelling the positive contribution from the seagull term—
a feature that has been demonstrated analytically within the FG
framework \cite{Cas25}.

An ingredient absent in single--particle models is the presence of
short--range correlations (SRC) in the nuclear wave function. Since
MEC are two--body operators, it is natural to expect them to be
sensitive to the short--distance structure of correlated nucleon
pairs, and therefore to investigate whether the enhancement of the
1p1h transverse response may be connected to such correlations.
Quantifying this effect is the main goal of the present work, which we
address within the simplest possible framework: the independent--pair
approximation in nuclear matter. Our motivation comes from the
calculation of Fabrocini \cite{Fab97} in the correlated basis function
(CBF) approach, where correlations were incorporated through explicit
operators in the wave function and MEC were included in the current
operator, leading to a significant enhancement of the 1p1h transverse
response.  Here, we adopt a completely different approach, in which
the wave functions of correlated pairs are obtained by solving the
Bethe--Goldstone (BG) equation. We show that the resulting high--momentum
components of the correlated pairs naturally produce a transverse
enhancement.  We employ the realistic Granada 2013 nucleon--nucleon
potential, obtained from a partial-wave analysis of nucleon-nucleon
scattering below pion production threshold
\cite{Nav13a,Nav13b}, which allows for an efficient solution of the
BG equation \cite{Cas23c}. This makes it possible to
incorporate high-momentum components into the Fermi gas response
formalism with minimal numerical cost. In this work, we briefly
present the formalism and show that the contribution of each partial
wave of the correlated wave function can be clearly identified.  We
find that the enhancement is predominantly produced by the
$^3S_1$--$^3D_1$ partial wave, in agreement with Fabrocini’s
conclusion that tensor correlations drive the enhancement, a mechanism
also suggested earlier by Leidemann and Orlandini \cite{Fab97,Lei90}.

\section{Formalism}

\subsection{Uncorrelated response function}

The starting point of this work are the longitudinal and transverse
response functions for quasielastic electron scattering,
defined as
\begin{equation}
  R_L(q,\omega)= W^{00},     \kern 1cm
  R_T(q,\omega)=W^{11}+W^{22},
  \label{eq:responses}
\end{equation}
where $q$ and $\omega$ are the momentum and energy transfer,
respectively, and \( W^{\mu\nu} \) is the inclusive hadronic tensor,
defined below.  This definition assumes that the $z$-axis points to
the momentum transfer $\nq$. Only the diagonal elements of the
hadronic tensor are involved.  We employ the non-relativistic
Fermi gas model, which provides the simplest framework to clearly
identify the effect under study without additional complications.  The
single particle states are plane waves, and the ground state,
$|F\rangle$, is a Slater determinant with all the levels filled up to
the Fermi momentum $k_F$.  In this framework we restrict ourselves to
one-particle one-hole (1p1h) excitations.  The diagonal elements of
the hadronic tensor are then
\begin{eqnarray}
W^{\mu\mu}&=& \sum_{ph}
\left| \langle ph^{-1} | \hat{J}^{\mu}(\nq) | F \rangle \right|^2
\nonumber\\ 
&\times& \delta(E_{p}-E_{h}-\omega)
\theta(p-k_F)\theta(k_F-h).
\label{hadronic}
\end{eqnarray}
Here, the index $p$ stands for the set of quantum numbers 
$(\np, s_p, t_p)$, namely momentum, spin, and isospin, and similarly for $h$; 
therefore, the sums over $p$ and $h$ in the hadronic tensor  
involve sums over the three indices.

The transverse response involves only the $\mu=1,2$ components
of the electromagnetic current that is the sum of one-body (1b) plus
two-body (2b) currents (MEC) 
\begin{equation}
\hat{\nJ}(\nq) = 
\hat{\nJ}_{1b}(\nq) 
+\hat{\nJ}_{2b}(\nq),
\end{equation}
The matrix elements of these operators between the ground state and 
a 1p1h excitation are given by
\begin{equation}
\left\langle ph^{-1} \right| \hat{J}^\mu_{1b} |\left. F \right\rangle
=
\left\langle [p] \right|\hat{J}^\mu_{1b} |\left. [h] \right\rangle,
\label{melement1}
\end{equation}
\begin{equation}
\left\langle ph^{-1} \right|\hat{J}_{2b}^{\mu} |\left. F \right\rangle 
=
\sum_{k<k_F}
\left\langle [pk] \right|\hat{J}_{2b}^{\mu} |\left. [hk]-[kh] \right\rangle 
\label{melement}
\end{equation}
where, for convenience, we introduce the notation $|[p]\rangle$ (with
brackets) to denote states normalized to unity in a finite volume $V$,
i.e. with spatial wave function $e^{i\np\cdot\nr}/\sqrt{V}$.  In
contrast, states without brackets, $|p\rangle$, refer to continuum
states with spatial wave function $e^{i\np\cdot\nr}/(2\pi)^{3/2}$,
normalized to a momentum delta function in the whole space.

We see in Eq (\ref{melement})
that the 2b current is acting over all the pairs $|hk\rangle$,
where $k$ is occupied.  The nucleon $|k\rangle = |ks_kt_k\rangle$ acts
as an spectator nucleon that does not change its state.

The  elementary matrix elements of the 1b and 2b currents
between plane waves states can be written as:
 \begin{eqnarray}
 \langle [p] |\hat{J}_{1b}^{\mu} | [h]\rangle 
&=&
  \frac{(2\pi)^{3}}{V}\delta^{3}(\nq+\nh-\np)
j_{1b}^{\mu}(p,h), 
\label{OBmatrix}
\\
\langle [p'_{1}p'_{2}]|\hat{J}_{2b}^{\mu}|[p_{1}p_{2}]\rangle
&=&
\frac{(2\pi)^{3}}{V^{2}}\delta^{3}(\np_1+\np_2+\nq-\np'_1-\np'_2)
\nonumber\\
&&
{}\times
j_{2b}^{\mu}(p'_1,p'_2,p_1,p_2) , 
\label{TBmatrix}
\end{eqnarray}
 where the Dirac deltas arise
 from momentum conservation.  The current functions $j^\mu_{1b}(p,h)$
 and \(j_{2b}^{\mu}(p'_1,p'_2,p_1,p_2)\) implicity depend on spin and
 isospin indices.
 Then we can write
  \begin{eqnarray}
\left\langle ph^{-1} \right| \hat{J}^\mu_{1b} |\left. F \right\rangle
&=&
  \frac{(2\pi)^{3}}{V}\delta^{3}(\nq+\nh-\np)
  j_{1b}^{\mu}(p,h),
  \label{j1b}
\\
\left\langle ph^{-1} \right| \hat{J}^\mu_{2b} |\left. F \right\rangle
&=&
  \frac{(2\pi)^{3}}{V}\delta^{3}(\nq+\nh-\np)
j_{2b}^{\mu}(p,h), 
\label{j2b}
  \end{eqnarray}
where the current function $j^\mu_{2b}(p,h)$, corresponding to the 2b
matrix element (\ref{melement}), is given by
\begin{eqnarray}
j_{2b}^{\mu}(p,h) 
\equiv 
\frac{1}{V}
\sum_{k<k_F}
\left[ j_{2b}^{\mu}(p,k,h,k)-j_{2b}^{\mu}(p,k,k,h)\right]
\label{effectiveOB}
\end{eqnarray}

By inserting Eqs. (\ref{j1b},\ref{j2b}),
into Eq. (\ref{hadronic}), taking the
thermodynamic limit $\sum_h \rightarrow V \sum_{s_h t_h}\int d^3h/(2\pi)^3$, 
and integrating over $\np$,
 the hadronic tensor can be written as
\begin{eqnarray}
W^{\mu\mu}
&=&
\frac{V}{(2\pi)^{3}}
\sum_{t_h}
\int d^3h\delta(E_{p}-E_{h}-\omega)
2 w^{\mu\mu}(p,h) \nonumber \\
&\times&
\theta(p-k_{F})\theta(k_{F}-h),
\label{integralw}
\end{eqnarray}  
where  \(\np = \nh + \nq\), and $t_p=t_h$.
The  function $w^{\mu\nu}$ is the effective single-nucleon hadronic tensor 
defined by 
\begin{eqnarray}
w^{\mu\mu}(p,h)
&=& \frac{1}{2}\sum_{s_ps_h} |j^\mu_{1b}(p,h)+j^\mu_{2b}(p,h)|^2
\label{single-nucleon}
\end{eqnarray}
Note that the single-nucleon tensor implicitly depends on the isospin
projection, and the sum over $t_h$ in Eq.~(\ref{integralw}) accounts
for both proton and neutron emission. The effective single-nucleon
tensor also includes the contribution of MEC, which leads to
interference between one-body and two-body currents. Indeed, by
expanding the square in Eq.~(\ref{single-nucleon}),
\begin{eqnarray}
w^{\mu\mu}(p,h)
&=&
  w^{\mu\mu}_{1b}+ w^{\mu\mu}_{1b2b}+w^{\mu\mu}_{2b},
\end{eqnarray}
where 
\begin{eqnarray}
  w^{\mu\mu}_{1b}
   & = & \frac{1}{2}\sum_{s_ps_h}
  |j^\mu_{1b}(p,h)|^2, \\
  w^{\mu\mu}_{1b2b} & = & \mbox{Re}\sum_{s_ps_h}
  j^\mu_{1b}(p,h)^*j^\mu_{2b}(p,h), \label{w1b2b}\\
w^{\mu\mu}_{2b}   & = & \frac{1}{2}\sum_{s_ps_h} |j^\mu_{2b}(p,h)|^2.
\end{eqnarray}
The first term, $w^{\mu\mu}_{1b}$, corresponds to the single-nucleon
response induced by the one-body current, i.e., the impulse
approximation, which produces the $1p1h$ response in this
approximation. The second term, $w^{\mu\mu}_{1b2b}$, represents the
genuine interference between one-body and two-body current
contributions, and it is the main focus of this work, since it
provides the largest MEC contribution to the $1p1h$ transverse
response. Finally, the third term, $w^{\mu\mu}_{2b}$, corresponds to
the pure MEC contribution, which is disregarded here because its
effect is small, as shown in ref. \cite{Cas23b}.  
Consequently, we write the total transverse response
as the sum of the $1b$ response and the 1b-2b interference response.
\begin{equation}
R^T= R^T_{1b}+ R^T_{1b2b}
\end{equation} 

The 1b current function used here is
\begin{eqnarray}
  \nj_{1b}(\np,\nh)&=&
-\delta_{t_pt_h}\frac{G_M^h}{2m_N}i\nq\times\nsigma_{s_ps_h},
\label{magnetization}
\end{eqnarray}
Here $G_M^h$ is the magnetic form factor of the nucleon with isospin
$t_h$.  In the quasielastic peak the convection current contribution
is much smaller that the magnetization current and can be neglected.
See, for instance, Ref. \cite{Cas25},
where a comparison between the interference of the convection and
magnetization currents with MEC is performed.

The two-body current in this work
consists of pion-exchange contributions, written as the sum of
seagull, pion-in-flight, and $\Delta$-excitation terms. These currents
are treated in the non-relativistic limit and correspond to the
standard MEC that are commonly considered as the leading exchange
contributions \cite{Ris79,Fru84,Eri88,Ris89,Sch89}.
Written as operators in spin-isospin space they are given by 
\begin{widetext}
\begin{eqnarray}
\nj_{s}(p'_1,p'_2,p_1,p_2)
&=&
i[\ntau^{(1)} \times \ntau^{(2)}]_z 
\frac{f^{2}}{m_{\pi}^{2}}F_1^V
\left(
\frac{\nk_1\cdot\nsigma^{(1)}}{\nk_1^2+m_{\pi}^2}
\nsigma^{(2)}
-\frac{\nk_2\cdot\nsigma^{(2)}}{\nk_2^2+m_{\pi}^2}
\nsigma^{(1)}
\right).
\label{seagull}
\\
\nj_{\pi}(p'_1,p'_2,p_1,p_2)
&=&
i[\ntau^{(1)} \times \ntau^{(2)}]_z
\frac{f^{2}}{m_{\pi}^{2}}F_1^V
\frac{\nk_1\cdot\nsigma^{(1)}}{\nk_1^2+m_{\pi}^2}
\frac{\nk_2\cdot\nsigma^{(2)}}{\nk_2^2+m_{\pi}^2}
(\nk_1-\nk_2).
\label{pionic}
\\
\nj_\Delta(p'_1,p'_2,p_1,p_2)
&=&
i \sqrt{ \frac32 } \frac29 
\frac{ff^*}{m_\pi^2}
\frac{C_3^V}{m_N}\frac{1}{m_\Delta-m_N}
\left\{
\frac{\nk_2\cdot\nsigma^{(2)}}{\nk_2^2+m_{\pi}^2}
\left[
4\tau_{z}^{(2)}\nk_2+
[\tau^{(1)}\times\tau^{(2)}]_{z}
\nk_2\times\nsigma^{(1)}
\right]
\right.
\nonumber\\
&&
\kern 4cm 
\left. \mbox{}+
\frac{\nk_1\cdot\nsigma^{(1)}}{\nk_1^2+m_{\pi}^2}
\left[
4\tau_{z}^{(1)}\nk_1-
[\tau^{(1)}\times\tau^{(2)}]_{z}
\nk_1\times\nsigma^{(2)}
\right]
\right\}
\times\nq.
\label{delta}
\end{eqnarray} 
\end{widetext}
with $\nk_1=\np'_1-\np_1$ and $\nk_2=\np'_2-\np_2$, and
$\nk_1+\nk_2=\nq$.  These MEC operators involve several standard
parameters: $f=1$ is the $\pi N$ coupling constant, $f^{*}=2.13$ is
the $\pi N\Delta$ coupling, $\boldsymbol{\tau}^{(i)}$ denotes the
isospin operator of nucleon $i$, $F_1^V=F_1^p-F_1^n$ is the isovector
nucleon form factor, $C_3^V$ is the $\Delta$ vector form factor,
$m_\pi$ is the pion mass, and $m_\Delta$ is the $\Delta$ mass. Further
details can be found in Ref.~\cite{Cas25}.  The values of the
couplings and form factors may vary in other works, and in many cases
a strong form factor is introduced at the $\pi NN$ and $\pi N\Delta$
vertices. 
The impact of these form factors is minor in the case of
the uncorrelated Fermi gas \cite{Cas25} because the momentum transfer to each
nucleon, $\nk_i$, is small for low $q$. However, in the correlated
contribution their impact becomes important because in this case $\nk_i$
depends on the high-momentum components and is not small. Therefore,
the strong form factors have been included in the present calculation.


The calculation of the interference response in the Fermi gas model
has been recently reviewed in detail in Ref.~\cite{Cas25} and will not
be repeated here; the result is that the total interference is
predominantly negative, driven by the destructive interference with
the $\Delta$ current. Moreover, Ref.~\cite{Cas25} showed that several
single-particle models yield qualitatively similar results for the
1b--2b transverse interference: although quantitative differences
exist, all approaches consistently predict a negative total
interference.
Hence, no enhancement arises unless an additional ingredient, absent
from all these models, is included---namely, correlations between
nucleons. SRC provide such a mechanism by combining MEC with the
high-momentum components of correlated pairs, yielding the required
enhancement.

\subsection{Short-range correlations}

In order to incorporate correlations into our formalism in a simple and
transparent way, we invoke the independent-pair
approximation, where nucleon pairs interact within the nuclear medium.
Recall that in Eq.~(5) the exchanged photon is absorbed by a pair of
non-interacting nucleons described by the plane-wave state
$|hk\rangle$. Once the nucleon--nucleon interaction is switched on, the
wave function of the pair is modified, acquiring high-momentum
components that reflect the presence of short-range correlations:
In practice, we replace the uncorrelated plane-wave state $|hk\rangle$
by the correlated wave function
\begin{equation}
|\Psi_{hk}\rangle = |ph\rangle + |\Phi_{hk}\rangle,
\end{equation}
where  $|\Phi_{hk}\rangle$ represents the high-momentum
component. The latter acquires strength above the Fermi momentum
because, in the nuclear medium, low-momentum states are Pauli blocked
and ordinary scattering between $h$ and $k$ cannot occur.  In order to
construct the correlated two--nucleon wave function
$|\Psi_{hk}\rangle$, we require that the two nucleons interact through
the NN potential, $V$, and that the wave function satisfies a
Schrödinger equation with Pauli blocking of intermediate states below
the Fermi momentum. This condition is expressed by the
Bethe--Goldstone equation \cite{Bet57,Gol57}
\begin{equation}
|\Psi_{hk}\rangle= |ph\rangle + \frac{Q}{E-T}V|\Psi_{hk}\rangle,
\end{equation}
where $E$ is the energy of the pair, $T$ is the kinetic energy and $Q$
is the projector over momenta above $k_F$.

Since the NN potential conserves the total momentum of the
two--nucleon system, it is convenient to switch from single--particle
coordinates $\nr_1,\nr_2$ to relative, $\nr=\nr_1-\nr_2$, and
center--of--mass, $\nR=(\nr_1+\nr_2)/2$, coordinates. Correspondingly,
the momenta are expressed in terms of the total momentum, $\mathbf{P}
= \mathbf{h}+\mathbf{k}$, and the relative momentum,
$\np=(\nh-\nk)/2$, the wave functions are written
\begin{eqnarray}
  \langle \nr_1\nr_2  |hk\rangle
  &=& \langle \nR|\nP\rangle \langle \nr |\np;s_p s_h\rangle
  \\
  \langle\nr_1\nr_2 |\Psi_{hk}\rangle
  &=& \langle \nR|\nP\rangle \langle \nr|\psi_{\nh\nk}^{s_ps_h}\rangle
  \end{eqnarray}
Here $|\mathbf{p};s_hs_k\rangle$ denotes the relative wave function of
the uncorrelated pair of nucleons with spin projections $s_h$ and
$s_k$, while $|\psi_{\nh\nk}^{\,s_hs_k}\rangle$ represents the
corresponding correlated relative wave function.
Note that the center--of--mass wave function $|\nP\rangle$
remains unaffected by correlations.

Since the NN potential commutes with $S^2$ and therefore conserves the
total spin $S=0,1$ of the two nucleons, it is convenient to express both
the uncorrelated and correlated relative states in the coupled spin
basis
\begin{eqnarray}
|\np; s_h s_k\rangle
&=& \sum_{SM_S}
\langle \tfrac{1}{2}s_h\tfrac{1}{2}s_k | S M_S \rangle\;
|\np; S M_S\rangle ,
\\
|\psi_{\nh\nk}^{\,s_h s_k}\rangle
&=& \sum_{SM_S}
\langle \tfrac{1}{2}\,s_h\;\tfrac{1}{2}\,s_k \,|\, S M_S \rangle\;
|\psi^{SM_S}_{\nh\nk}\rangle .
\end{eqnarray}
The state $\psi_{\nh\nk}^{SM_S}$ denotes the relative correlated wave
function obtained from the uncorrelated relative state
$|\np;SM_S\rangle$,
where the two nucleons are coupled to total spin
$S$ with projection $M_S$.
Exploiting the conservation of both the total momentum $\nP$ and the
total spin $S$, the BG equation can be formulated for
the relative correlated wave function $|\psi_{\nh\nk}^{SM_S}\rangle$.
\begin{eqnarray}
  |\psi_{\nh\nk}^{SM_S}\rangle
  &=& |\np;SM_S\rangle
+\int d^3p'\frac{\overline{Q}(P,p')}{p^2-p'{}^2}
\nonumber\\
&& \kern -1cm \mbox{}\times
\sum_{M'_S}|\np';SM'_S\rangle
\langle \np';SM'_S| m_NV  |\psi_{\nh\nk}^{SM_S}\rangle
\end{eqnarray}
Since the nucleon--nucleon potential does not conserve the value of
the spin projection $M_S$, the correlated wave function for $S=1$ in
general contains admixtures of all three components $M_S=-1,0,1$.
Here $\overline{Q}(P,p)$ denotes the Pauli projector averaged over the
angles of the center--of--mass momentum. The use of the
angle--averaged projector ensures that the correlated wave functions
depend only on the modulus of the total momentum, thereby simplifying
the solution of the BG equation in a partial--wave
expansion. This approximation is standard and accurate in nuclear
matter calculations~\cite{Preston-Bhaduri}, and it is essential to
make the calculation numerically feasible.

We do not describe here the details of the resolution of the
Bethe--Goldstone equation in the partial--wave expansion. The
interested reader can find the complete discussion in
Ref.~\cite{Cas23c}, in the particular case of the realistic Granada
2013 potential. This potential is a coarse-grained interaction
represented as a sum of delta functions in each partial wave, which
greatly simplifies the solution of the BG equation, reducing it to a
straightforward matrix inversion. More details can be found in
Ref.~\cite{Cas23c}.

The effect of
correlations on the interference response is linear in the defect function,
i.e., the high--momentum component of the relative wave function
defined as
\begin{equation}
  |\phi_{\nh\nk}^{SM_S}\rangle = |\psi_{\nh\nk}^{SM_S}\rangle -
|\np,SM_S\rangle.
\end{equation}
It is computed as a sum of multipoles. For this work we need the wave
function in momentum space, given as
\begin{eqnarray}
  \phi_{\nh\nk}^{SM_S}(\np')
  &=&\sqrt{\tfrac{2}{\pi}}
  \sum_{JM}\sum_{ll'm}i^{l'-l}Y_{l'm}(\hat{\np})
  \langle l'mSM_S|JM\rangle
  \nonumber\\
  &&\times
  \phi^{SJ}_{ll'}(P,p;p'){\cal Y}_{lSJM}(\hat{\np}')
\end{eqnarray}
The defect partial waves
$\phi_{ll'}^{SJ}(P,p;p')$, which are functions of the initial and
final (high) relative momenta $p,p'$ of the pair, and also of the CM momentum
$P$, are obtained from the solution of the BG equation
for each partial wave.
These functions constitute the input of the
present calculation of the response function.
In practice, only the
lowest partial waves of the expansion are needed, since the
interacting nucleons have energies below the Fermi surface. The most
relevant contributions correspond to the $^1S_0$ and $^3S_1$--$^3D_1$
channels. The angular-spin dependence of the defect function
is expressed as a linear
combination of the bi--spinorial spherical harmonics ${\cal
  Y}_{lSJM}(\hat{\np}')=\sum_{lM_S}\langle lmSM_S|JM\rangle
Y_{lm}(\hat{\np}')|SM_S\rangle$, where $\hat{\np}'$ denotes the angles
of the
relative high-momentum of the correlated pair.  As we shall show below, the
transverse enhancement arises predominantly from the $^3S_1$--$^3D_1$
channel, corresponding to $S=J=1$ and $l,l'=0.2$.

\begin{figure*}[ht]
  \centering
\includegraphics[width=13cm,bb=100 460 520 690]{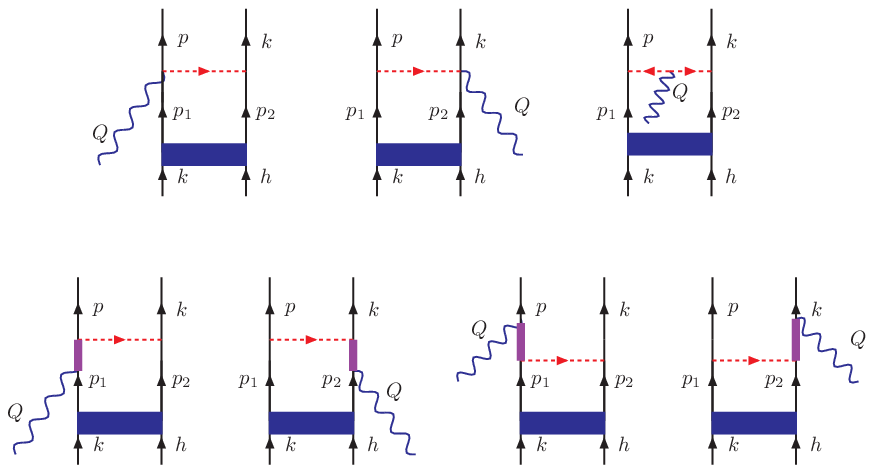}
  \caption{Feynman diagrams illustrating the action of the two-body
    MEC on the high-momentum components of the correlated nucleon
    pair. The exchange contributions of the seagull and  pionic (top), and
    $\Delta$ (bottom) currents are shown, where the photon is absorbed by the
    high-momentum intermediate legs generated by the correlation
    between nucleons $h$ and $k$.}
\label{fig1}
\end{figure*}

\subsection{1p1h MEC matrix element with SRC}

Within the independent--pair approximation, the uncorrelated two--nucleon
state $|hk\rangle$ is replaced by the \emph{correlated} state
$|\Psi_{hk}\rangle$ obtained from the BG equation.
In the calculation of the MEC matrix elements, this substitution implies
that the exchanged photon can also couple to the
high--momentum components generated by the interaction of the nucleon pairs.
Accordingly, the 1p1h MEC matrix element is modified as
\begin{eqnarray}
\label{MECcorr}
\kern -1cm \langle ph^{-1}|\hat{J}^{\mu}_{2b}|F\rangle
&\longrightarrow&
\sum_{k<k_F}
\langle [p\,k]|\hat{J}^{\mu}_{2b}|[\Psi_{hk}-\Psi_{kh}]\rangle
\nonumber\\
&& \kern -1cm \mbox{}=
\langle ph^{-1}|\hat{J}^{\mu}_{2b}|F\rangle_{\rm FG}+
\langle ph^{-1}|\hat{J}^{\mu}_{2b}|F\rangle_{\rm cor}
\end{eqnarray}
where the subscript ``FG'' denotes the uncorrelated matrix element in
the Fermi gas, while the subscript ``cor'' refers to the additional correlated
contribution arising from the high--momentum components of the nucleon
pairs, i.e., from the defect functions.
\begin{equation}
\langle ph^{-1}|\hat{J}^{\mu}_{2b}|F\rangle_{\rm cor}
=
\sum_{k<k_F}
\langle [p\,k]|\hat{J}^{\mu}_{2b}|[\Phi_{hk}-\Phi_{kh}]\rangle
\end{equation}
In order to compute the correlated MEC matrix element it is necessary
to first perform the isospin sum. For this purpose, it is convenient
to separate explicitly the two--body isospin structure from the
current operators. From Eqs. (\ref{seagull},\ref{pionic},\ref{delta}),
the full two-body current operator
is a linear combination of three
basic isospin operators,
$U_1=  \tau_{z}^{(1)}$,
$U_2=\tau_{z}^{(2)}$,
$U_3=i[\ntau^{(1)}\times\ntau^{(2)}]_{z}$,
reflecting their purely
isovector nature and the conservation of charge,
\begin{equation}
\hat{J}^\mu_{2b}=
  \tau_{z}^{(1)}J_1^\mu+
\tau_{z}^{(2)}J_2^\mu+
i[\ntau^{(1)}\times\ntau^{(2)}]_{z}J_3^\mu
  \equiv \sum_{a=1}^3 U_a J_a^\mu,
\end{equation} 
where the 2b operators  $J_a^\mu$ depend only on the momenta
and spins of the two nucleons.

The explicit evaluation of the isospin matrix elements and sum over
$t_K$,
is deferred to the Appendix.  The result can be written as
\begin{eqnarray}
  \sum_{t_k}  \langle pk| \hat{J}_{2b}^\mu | \Phi_{hk}-\Phi_{kh}\rangle
  &=&
 2t_h\delta_{t_pt_h} \langle pk| J_{\rm dir} | \Phi_{hk}\rangle
\nonumber\\
&&\kern -2cm
+ 2t_h\delta_{t_pt_h} \langle pk| J_{\rm exch} | \Phi_{kh}\rangle
\end{eqnarray}
where we the effective currents appearing in the direct and exchange
matrix elements are
\begin{equation}
  J_{\rm dir}^\mu=2J_1^\mu; \kern 1cm J_{\rm exch}=2J_3^\mu-J_1^\mu-J_2^\mu
\end{equation}

Finally, in the Appendix we derive the correlated contribution to the
1p1h MEC matrix element, which accounts for the high-momentum
components of the pair wave function, and show that it can be
expressed  as an integral over the relative
defect function in momentum space, yielding a compact final form:
\begin{equation}
\left\langle ph^{-1} \right| \hat{J}^\mu_{2b} |\left. F \right\rangle_{\rm cor}
=
  \frac{(2\pi)^{3}}{V}\delta^{3}(\nq+\nh-\np)\,
j_{\rm cor}^{\mu}(p,h),
\end{equation}
with
\begin{equation}
j_{\rm cor}^\mu(p,h)= 2t_h \delta_{t_pt_h}
[j_{\rm dir}^\mu(\np s_p,\nh s_h)+ j_{\rm exch}^\mu(\np s_p,\nh s_h)],
\end{equation}
where the direct and exchange contributions are explicitly given by
\begin{widetext}
\begin{eqnarray}
j_{\rm dir}^{\mu}(\np s_p,\nh s_h) &=& 
\frac{1}{V}\sum_{\nk}\sum_{s_k}\sum_{r_1r_2}
  \int \dif p_1 \,
  j_{\rm dir}^\mu(\np s_p,\nk s_k; \np_1 r_1, \np_2 r_2)\,
  \phi_{\nh\nk}^{s_hs_k}\left(\frac{\np_1-\np_2}{2}\right)_{r_1r_2},
  \\
j_{\rm exch}^{\mu}(\np s_p,\nh s_h) &=& 
\frac{1}{V}\sum_{\nk}\sum_{s_k}\sum_{r_1r_2}
  \int \dif p_1 \,
  j_{\rm exch}^\mu(\np s_p,\nk s_k; \np_1 r_1, \np_2 r_2)\,
  \phi_{\nk\nh}^{s_ks_h}\left(\frac{\np_1-\np_2}{2}\right)_{r_1r_2}
\end{eqnarray}
\end{widetext}
The
integration over $\np_1$, with $\np_2=\nh+\nk-\np_1$, is restricted to
the high--momentum region $p_1,p_2>k_F$, since the defect function
$  \phi_{\nk\nh}^{s_ks_h}$
vanishes by definition when these momenta are Pauli blocked. Thus, the
two--body current acts effectively only on the high--momentum
components of the correlated pair wave function, which is precisely
the mechanism by which SRC enter the interference contribution.

The effect of this correlation current is illustrated in the Feynman
diagrams of Fig.~1, where we show the seagull, pionic, and $\Delta$
diagrams for the exchange part. We see that the photon is attached to
the high-momentum intermediate legs that result from the correlation
between $h$ and $k$. The corresponding direct diagrams have the same
structure, but with the roles of initial $h$ and $k$ interchanged.

\begin{figure*}[ht]
  \centering
\includegraphics[width=13cm,bb=64 458 492 805]{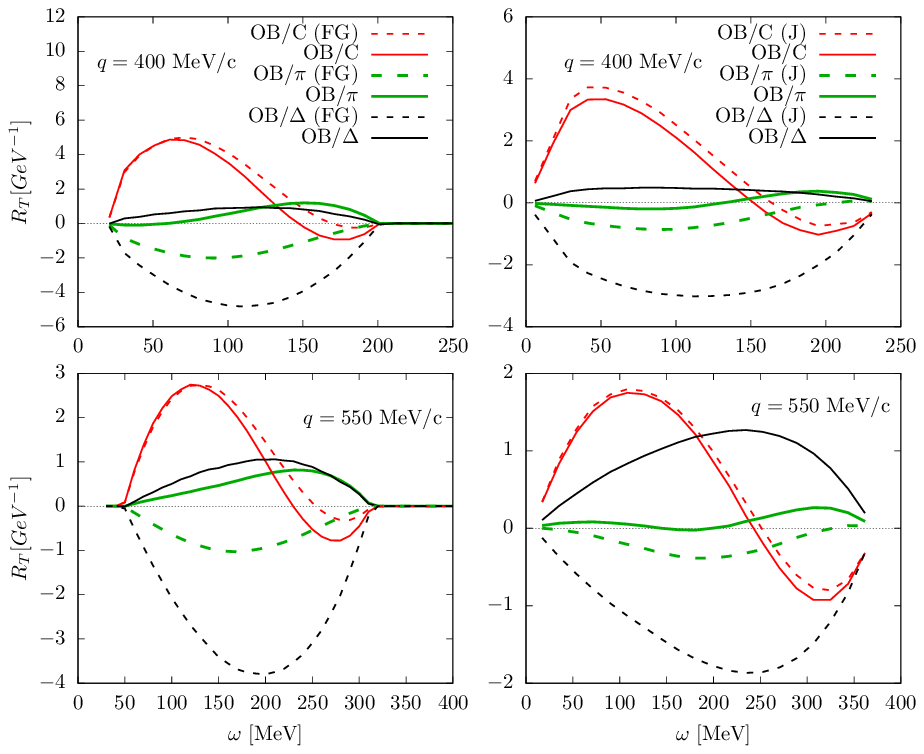}
\caption{Comparison of the transverse 1p1h interference responses,
  obtained in the present calculation (left panels) for $^{12}$C, with
  those from the correlated basis function (CBF) calculation of
  Fabrocini~\cite{Fab97} (right panels). The separate contributions
  from seagull (OB/C, red lines), pionic (OB/$\pi$, green lines), and
  $\Delta$ (OB/$\Delta$, black lines) currents are shown for momentum
  transfers $q=400$~MeV/c and $q=550$~MeV/c.  In the left panels we
  show the results without (FG) correlations and total.  In the right
  panels we show the results with Jastrow correlations only (J) and
  the total.  }
\label{fig2}
\end{figure*}

\section{Results}

Having established the theoretical framework showing how SRC modify
the 1p1h MEC matrix elements within the independent--pair
approximation, through their coupling to the high--momentum components
of nucleon pairs in the medium, we now turn to the computation of
their contribution to the interference between one--body and two--body
currents in the transverse response. Specifically, the SRC correction
to the transverse response is added to the previously calculated
uncorrelated MEC  result of ref. \cite{Cas25}.
The new term takes the same structure
as Eq.~(15), with the replacement of the two--body current by its
correlated part, namely
\begin{equation}
w^{\mu\mu}_{1b2b,\rm cor} = \mbox{Re}\sum_{s_p s_h}
  j^\mu_{1b}(p,h)^*\, j^\mu_{\rm cor}(p,h).
\end{equation}
Taking the one--body current of Eq.~(18) and performing the isospin
sums, one finds that the correlation contribution to the transverse
response, after summing over $t_p$ and $t_h$, takes the form
\begin{eqnarray}
  \sum_{\mu=1}^2 w^{\mu\mu}_{1b2b,\rm cor}
  &=&
  \mbox{Re}\sum_{s_p s_h}
  \frac{G_M^p-G_M^n}{2m_N}(i\nq\times\nsigma_{s_hs_p})^\mu 
  \nonumber\\
  &&
\kern -1cm \times [j^\mu_{\rm dir}(\np s_p,\nh s_h)
+j^\mu_{\rm exch}(\np s_p,\nh s_h)].
\end{eqnarray}

When
this single--nucleon amplitude  is inserted in the total
response function, Eq.~(11), the integration over $d^3h$ simplifies to
a one--dimensional integral in the hole momentum $h$. This reduction
occurs because the energy--conserving delta function fixes the polar
angle $\theta_h$, while axial symmetry around the momentum transfer
direction allows us to set $\phi_h=0$ and recover the full azimuthal
phase space through a multiplicative factor of $2\pi$.
In addition,  the calculation of the correlated matrix element
involves:
\begin{enumerate}
\item a three--dimensional integral over the spectator nucleon
  momentum $\mathbf{k}$, in the thermodinamic limit,
\begin{equation}
\frac{1}{V}\sum_{\nk}\rightarrow \int\frac{d^3k}{(2\pi)^3},
\end{equation}
restricted to values below the Fermi surface;
\item a second three--dimensional integral over the high--momentum
  component $\mathbf{p}_1$, which encodes the short--range correlation
  effects through the defect function in momentum space.
\end{enumerate}
These two integrations combine with the reduced one--dimensional
integral over the hole momentum $h$, yielding a total of seven
dimensional  numerical integrations that must be carried out for each
kinematical point $(q,\omega)$.

The calculation procedure requires solving the Bethe--Goldstone (BG)
equation for each nucleon pair $\nh,\nk$ in order to obtain the
corresponding defect function $\phi_{hk}$ in a multipole expansion.
The use of the Granada 2013 potential greatly simplifies this task:
due to its coarse--grained representation, the BG equation can be
solved in a straightforward manner without numerical
complications. The details of the BG solution with the Granada
potential are provided in Ref.~\cite{Cas23c}.

Rigorously, the integral over the high momentum $\np_1$ extends from
$k_F$ to infinity. In practice, however, the high--momentum radial
functions $\phi_{ll'}^{SJ}$ decrease as a power of the relative
momentum \cite{Cas23c}, so it is sufficient to integrate up to
$|\np_1|\sim 800$~MeV or less to obtain convergence.

As a first step in our investigation, we have studied the validity of the frozen
approximation for the spectator nucleon in the calculation of the
correlated responses. In this approach, within the integral over
$\nk$, one simply sets $\nk=0$ (frozen). With this assumption the
integration reduces to a factor $4\pi k_F^3/3$. We have found that
this approximation is quite accurate, introducing only a $\sim 6\%$
variation in the correlated interference transverse response. This
represents a major computational advantage, since the problem is
reduced from a seven--dimensional to a four--dimensional integral, and
it is no longer necessary to solve the BG equation for
all $h,k$ pairs but only for $h$ and $k=0$, which are far
fewer. Mathematically, the approximation can be understood as a direct
application of the mean value theorem for integrals, where we assume
the optimal midpoint to be at $\nk=0$. Physically, it is also
reasonable: in Ref.~\cite{Cas23c} we analyzed the dependence of the
defect wave functions $\phi_{ll'}^{SJ}$ on the pair center--of--mass
momentum and found it to be very weak. Moreover, the dependence on the
initial relative momentum $(\nh-\nk)/2$ is also very smooth. This
behavior is connected to the universality of the high--momentum
distribution.

In the following, the results are obtained using the frozen--$k$
approximation, which provides a numerically efficient procedure while
preserving the accuracy of the calculation.

We now assess the reasonableness of our results by comparing them with
other calculations of the 1b--2b interference including
correlations. The only available work of this kind is that of
Fabrocini~\cite{Fab97}, who employed the correlated basis function (CBF)
model in nuclear matter. A deeper comparison is possible since
Fabrocini published the separate contributions of the seagull, pionic,
and $\Delta$ currents. This comparison is shown in Fig.~2, where we
display the transverse 1b--seagull (or OB/contact), 1b--pionic (or
OB/pi), and 1b--$\Delta$ (OB/Delta) interference responses of $^{12}$C at
$q=400$ and $550$ MeV/c.  The left panels show our uncorrelated Fermi
gas results and the total results including correlations, while the right
panels show Fabrocini’s results with central (Jastrow) correlations
and the total.
To compute the interference 1b–2b responses for $^{12}$C, we use $k_F = 225$
MeV/c and apply the relation between the number of particles and the
volume of the Fermi gas in the thermodynamic limit. Consequently, the
factor appearing in the hadronic tensor, Eq. (11), is
$V/(2\pi )^3 = Z /[(8/3)\pi k_F^3]$, with $Z = 6$ for $^{12}$C.
Fabrocini’s results were
given as response per particle. Therefore, in Fig. 2 they have been
multiplied by a factor of 12 to compare with $^{12}$C.

The first striking feature in Fig. 2 is the qualitative similarity of
the results. In the independent-pair approximation (the present
calculation), the effect of correlations on the seagull response is
very small, so that it practically does not change when correlations
are included. The seagull response is large and positive, becoming
slightly negative beyond the midpoint of the quasielastic
region. Correlations have a more noticeable effect on the pionic
response, which is negative in the absence of correlations and becomes
positive when correlations are included. The most significant effect
is observed in the interference with the $\Delta$ current: in the
uncorrelated case it is negative and dominates the three
contributions, while correlations render it positive. This effect is
particularly pronounced at $q=550$ MeV/c.

By examining the CBF results in the right panels of Fig.~2, we see
that the Fermi gas with Jastrow correlations is very similar to the
uncorrelated Fermi gas, as noted in Ref.~\cite{Fab97}. However, the
inclusion of the full correlated wave function produces a noticeable
effect, closely resembling the outcome of our calculation: the
$\Delta$ and pionic contributions become positive, while the seagull
contribution remains almost unchanged.

\begin{figure}[ht]
  \centering
  \includegraphics[width=6cm,bb=170 280 385 800]{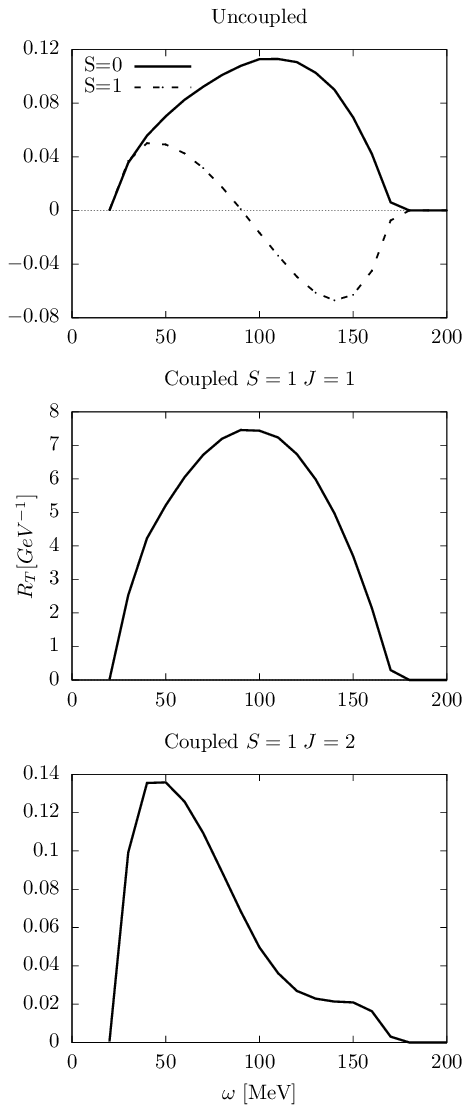}
\caption{Separate contributions to the interference transverse
  response of $^{12}$C for $q=380$ MeV/c, from different channels in
  the multipole expansion of the correlated wave function. Top panels:
  uncoupled channels with $S=0$ and $S=1$. Middle panels: coupled
  channels with $S=1$, $J=1$. Bottom panels: coupled channels with
  $S=1$, $J=2$. The dominant contribution clearly arises from the
  $^3S_1$--$^3D_1$ channel, showing the decisive role of tensor
  correlations in the nuclear medium.}
  \label{fig3}
\end{figure}

The qualitative similarity between our results and those of Fabrocini
indicates that the two models share compatible ways of describing
short-range correlations, which is remarkable given that the two
approaches treat correlations in fundamentally different ways. In the
CBF calculation, nuclear states are constructed by multiplying
uncorrelated Slater determinants by a correlation operator that
includes several types of correlations, such as spin, isospin, and
tensor combinations, with the correlation functions obtained
variationally using the Argonne v14 potential. In our approach, the NN
potential is different, the BG equation is solved, and
the independent-pair approximation is applied. Despite these very
different treatments of SRC, both approaches exhibit a similar
tendency in their effects on the interference 1b–2b transverse
response.

The different treatment of SRC may partly explain the quantitative
differences between both approaches; in addition, we note that the CBF
results were obtained with a somewhat larger value of $k_F$, which
also contributes to the discrepancy.  Finally, another reason for the
quantitative differences between the results may be that Fabrocini
modified the MEC operators to enforce gauge invariance with the
potential, as in the model of Schiavilla, Pandharipande and Riska
\cite{Sch89}, whereas in the present calculation we use the
stantard one-pion exchange currents. As a consequence, the
interference responses obtained in the Bethe--Goldstone calculation
are somewhat larger than those in the CBF model.

A deeper insight into the effect of correlations on the 1b--2b
interference response is presented in Fig.~3, where we show the
separate contributions in $^{12}$C, for $q=380$ MeV/c, from different
multipoles, $\phi_{ll'}^{SJ}$, of the correlated wave
function. Specifically, we display the contributions from the
uncoupled channels with $S=0$ and $S=1$ (top panels), the coupled
channels with $S=1$, $J=1$ (middle panels), and the coupled channels
with $S=1$, $J=2$ (bottom panels). It is evident that the dominant
contribution comes from the $S=1$, $J=1$ channels, i.e., the
$^3S_1$--$^3D_1$ components, which exceed the other contributions by
roughly a factor of 50. This dominance arises primarily from the
$^3S_1$--$^3D_1$ interference, indicating that the main effect is
driven by the tensor force in the NN interaction, a feature also
reported by Fabrocini in Ref.~\cite{Fab97}.  Fig. 3 indicates that
correlations tend to generate a significant enhancement of the
transverse response. Since the dominant contribution arises from the
$^3S_1$--$^3D_1$ channel, i.e. the deuteron channel, this enhancement
is produced by the interaction with correlated $np$ pairs in the
nuclear medium.

\begin{figure}[ht]
  \centering
\includegraphics[width=7cm,bb=170 285 385 800]{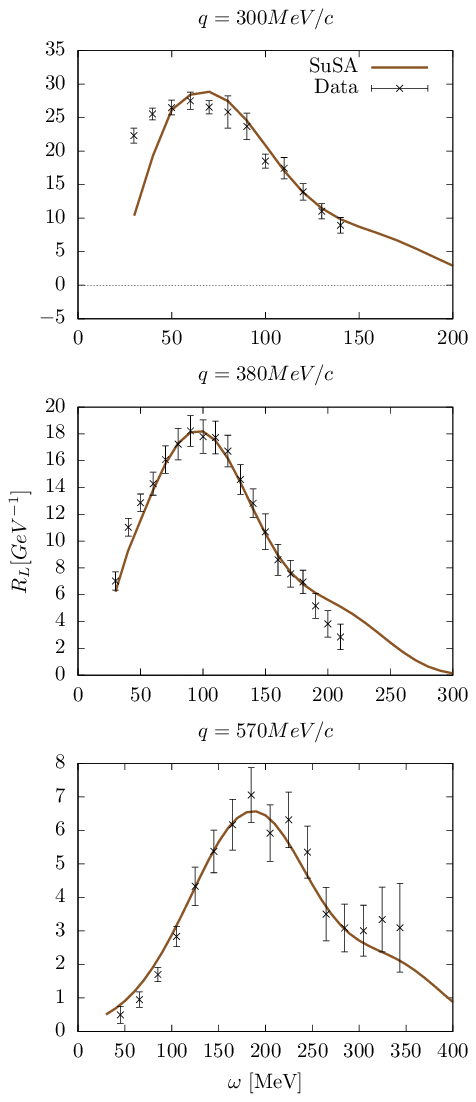}
\caption{(Color online) Longitudinal response function of $^{12}$C for
  three values of the momentum transfer $q$ within the SuSA model. The
  model assumes factorization of the response into the single-nucleon
  contribution times the scaling function $f_L(\psi')$. Data are taken
  from Ref.~\cite{Jou96,Bod22}.}  
\label{fig4}
\end{figure}

\begin{figure}[ht]
\centering
\includegraphics[width=7cm,bb=170 285 385 800]{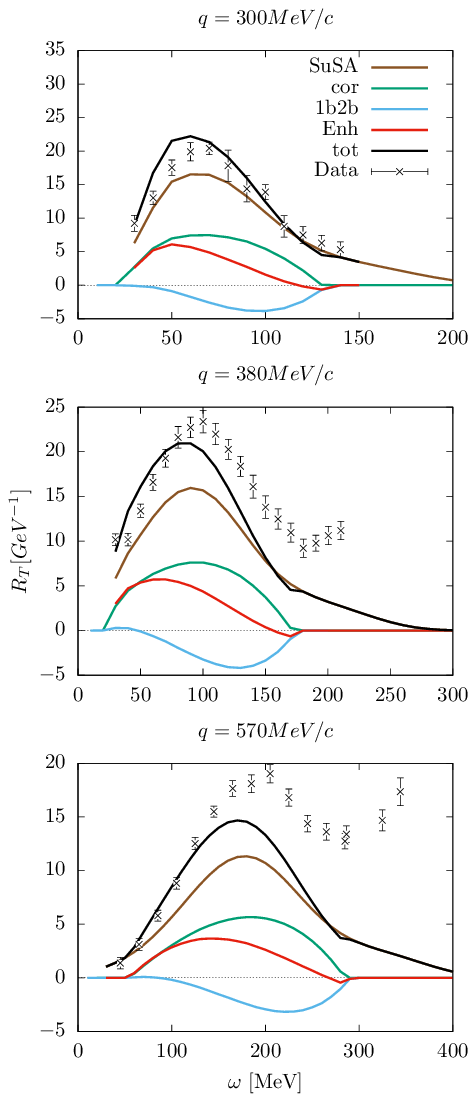}
\caption{Transverse response of $^{12}$C in the hybrid model for
  different momentum transfers. The one-body (1b) contribution from
  SuSA is shown, together with the uncorrelated FG 1b–2b interference
  and the total including correlations. The inclusion of the
  correlation contribution produces a positive enhancement, bringing
  the theoretical prediction closer to the experimental data.
  Data are taken   from Ref.~\cite{Jou96,Bod22}.}  
\label{fig5}
\end{figure}

A crucial test of our approach is provided by the comparison with
experimental data for the transverse response. This comparison is more
delicate, since the simple Fermi gas cannot directly reproduce the
longitudinal response and even less the transverse one without a
number of corrections. For this reason, we adopt here a hybrid scheme
in which the quasielastic one-body response is described using a
scaling model (SuSA), while the interference 1b–2b contribution is
computed in the Fermi gas and added incoherently to the SuSA one-body
response. The inherent uncertainty associated with the possible
inconsistency of this hybrid treatment is compensated by the
simplicity of the correlation model in the independent pair
approximation, which at least provides a clear indication of the
expected trend within a more sophisticated, though undoubtedly more
complex, approach that lies beyond the scope of this work.

In Fig.~4 we present the longitudinal response of $^{12}$C for three
 values of the momentum transfer within the SuSA
model. This approach assumes that the longitudinal response factorizes
into the product of the single-nucleon contribution and a universal
scaling function $f_L(\psi')$. The scaling variable $\psi'$ is defined
in terms of the minimum energy of a nucleon in the Fermi sea that can
absorb the transferred energy and momentum $(\omega,q)$. Further
details on the construction of the model can be found in
Ref.~\cite{Cas23a}. The comparison with data shows a very good
agreement, which reflects the well-established fact that the measured
longitudinal response exhibits scaling behavior to a high degree
within experimental uncertainties.

In Fig.~5 we present the predictions of the hybrid model for the
transverse response of $^{12}$C at the same values of $q$. The
quasielastic one-body response is evaluated with the SuSA model,
assuming $f_T(\psi')=f_L(\psi')$. As shown, the data lie well above
the one-body contribution, indicating that an additional enhancement
is required. The uncorrelated FG one-body–two-body interference is
negative, driving the response in the opposite direction of the
data. However, when correlations are included, the interference
response becomes positive and yields a net enhancement, which for
$q=300$ MeV/c is precisely of the magnitude needed to reproduce the
data. At higher momentum transfers an enhancement remains, although it
is insufficient to fully account for the data. This is expected, since
additional contributions from pion production and from the
two-particle–two-hole channel, which are not included here, become
increasingly important at larger $q$.

Finally, Fig. 5  highlights the significance of the
SRC contribution to the 1p1h MEC response.
Our approach is based on a microscopic calculation that starts from a
realistic NN interaction fitted to nucleon–nucleon scattering data,
without adjusting any additional parameters.
The BG equation is
solved for this potential, and the independent-pair approximation is
applied. Remarkably, this procedure produces an enhancement of the
transverse response that is precisely of the order of magnitude
required to explain the data. This result, which was not obvious
\emph{a priori}, demonstrates that the proposed SRC mechanism is
physically reasonable and underscores the practical relevance and
predictive power of this approach.

\section{Conclusions}

In this work we have investigated the effect of short-range
correlations on the 1p1h MEC contribution to the transverse response
of nuclei. Using the independent pair approximation, we have
incorporated SRC by replacing the uncorrelated pair wave function with
the correlated wave obtained from the solution of the Bethe–Goldstone
equation with a realistic NN interaction, namely the
Granada 2013 potential. This approach allows the two-nucleon wave
function to acquire high-momentum components above the Fermi sea,
which are absent in the uncorrelated Fermi gas.

The results clearly show that SRC generate a sizable enhancement of
the transverse 1b–2b interference response in the quasielastic peak
region. The multipole decomposition of the correlated wave function
reveals that the dominant contribution arises from the coupled $S=1$,
$J=1$ channel, i.e., the $^3S_1$–$^3D_1$ (deuteron-like)
components. This indicates that the enhancement is primarily driven by
the tensor component of the NN interaction, in agreement with previous
findings from CBF calculations.
Comparison with the CBF results shows a qualitative
agreement despite the very different treatments of SRC: while the CBF
approach applies a variational correlation operator to uncorrelated
Slater determinants, our method solves the BG equation.

Finally, we have combined our correlated 1b–2b response with a SuSA
description of the quasielastic 1b response in a hybrid model to
compare with experimental transverse response data of $^{12}$C. We
find that the inclusion of SRC produces a net enhancement of the
transverse response that brings the calculation into the correct order
of magnitude to describe the data at low and intermediate momentum
transfer. While at higher $q$ additional contributions from 2p2h
excitations and pion emission are missing, the present calculation
highlights the essential role of SRC in generating the transverse
enhancement.

Overall, this study demonstrates that the SRC mechanism incorporated
via the Bethe–Goldstone equation and the independent-pair
approximation provides a physically transparent and parameter-free
framework to account for the observed enhancement in the 1p1h MEC
transverse response.  The dominance of the $^3S_1$–$^3D_1$ channel
underscores the importance of the tensor force in the nuclear medium,
and the present results suggest that SRC and MEC should be included in
any realistic modeling of quasielastic transverse responses.

Future investigations stemming from this work include extending the
formalism to neutrino scattering, since the transverse enhancement
will appear in the vector response; it is also important to determine
whether the responses associated with the axial current exhibit a
similar enhancement due to SRC and MEC, which would have direct
implications for neutrino interaction models.

\section*{Acknowledgements}

The work was supported by Grant No.  PID2023-147072NB-I00 funded by
MICIU/AEI /10.13039/501100011033 and by ERDF/EU; by Grant No.  FQM-225
funded by Junta de Andalucia;

   \appendix

\begin{widetext}
   \section{Calculation of the correlated matrix element}

We assume that the correlated wave function $\Phi_{hk}$ is independent 
of the isospins $t_h$ and $t_k$, i.e., charge symmetry and charge 
independence are imposed. Although the Granada 2013 potential shows a 
small $np$--$pp$ difference in the ${}^1S_0$ wave, its effect on the 
high--momentum components is negligible.
Then the isospin matrix elements of the $U$-operators can be factorized.
Writing explicity the isospin indices, this means that, in the case of the direct matrix element
\begin{eqnarray}
\sum_{t_k}  \langle pk;t_pt_k| \hat{J}_{2b}^\mu | \Phi_{hk},t_ht_k\rangle
    = 
 \sum_a\left[
    \sum_{t_K}
  \langle t_pt_k|U_a | t_ht_k\rangle\right]
  \langle pk| J_a^\mu | \Phi_{hk}\rangle,
\end{eqnarray}
where the matrix element $\langle pk| J_a^\mu | \Phi_{hk}\rangle$ no
longer depends on isospin.
The basic sums over isospin for the direct and exchange terms are the following
\begin{eqnarray}
  \sum_{t_K}\langle t_pt_k|U_1 | t_ht_k\rangle = \delta_{t_pt_h}4t_h,
& \kern 1cm
  \sum_{t_K}\langle t_pt_k|U_2 | t_ht_k\rangle = 0,
&\kern 1cm
  \sum_{t_K}\langle t_pt_k|U_3 | t_ht_k\rangle = 0
  \\
  \sum_{t_K}\langle t_pt_k|U_1 | t_kt_h\rangle = \delta_{t_pt_h}2t_h
& \kern 1cm
  \sum_{t_K}\langle t_pt_k|U_2 | t_kt_h\rangle = \delta_{t_pt_h}2t_h
& \kern 1cm
  \sum_{t_K}\langle t_pt_k|U_3 | t_kt_h\rangle = -\delta_{t_pt_h}4t_h
\end{eqnarray}
By applying the isospin sums for the three operators, the total
isospin sum in the matrix element of the current can be written as
\begin{eqnarray}
  \sum_{t_k}  \langle pk| \hat{J}_{2b}^\mu | \Phi_{hk}-\Phi_{kh}\rangle
  =
 2t_h\delta_{t_pt_h} \langle pk| 2J_1^\mu | \Phi_{hk}\rangle
+ 2t_h\delta_{t_pt_h} \langle pk| 2J_3^\mu-J_1^\mu-J_2^\mu | \Phi_{kh}\rangle
\end{eqnarray}
At this stage, the terms 
$\langle pk|\,2J_1^\mu\,| \Phi_{hk}\rangle$ and 
$\langle pk|\,(2J_3^\mu-J_1^\mu-J_2^\mu)\,| \Phi_{hk}\rangle$ 
are purely spin--momentum matrix elements, since the isospin dependence has already been factorized and eliminated through the isospin sums.

To evaluate it, we insert a complete set of
momentum-spin states $|\np_1 r_1, \np_2 r_2\rangle$, making the spin dependence
explicit:
\begin{eqnarray}
  \langle [\np s_p,\nk s_k ]|J_a^\mu(\nq) |[\Phi_{\nh\nk}^{s_hs_k}]\rangle
 & =&
\int \dif p_1 \dif p_2 \sum_{r_1r_2}
\langle [\np s_p,\nk s_k ]|J_a^\mu(\nq)
|\np_1 r_1, \np_2 r_2\rangle\langle p_1 r_1, p_2 r_2|
[\Phi_{\nh\nk}^{s_hs_k}]\rangle
\nonumber\\
 & =&
\int \dif p_1 \dif p_2 \sum_{r_1r_2}
\langle [\np s_p,\nk s_k ]|J_a^\mu(\nq)
|[\np_1 r_1, \np_2 r_2]\rangle
\langle \np_1 r_1, \np_2 r_2|\Phi_{\nh\nk}^{s_hs_k}\rangle
\label{line2}\\
 &&
\kern -4cm =
\int \dif p_1 \dif p_2 \sum_{r_1r_2}
\frac{(2\pi)^3}{V^2}
\delta(\np_1+\np_2+\nq-\np-\nk)
  j_a^\mu(\np s_p,\nk s_k; \np_1 r_1, \np_2 r_2)
  \delta(\np_1+\np_2-\nh-\nk)
  \phi_{\nh\nk}^{s_hs_k}(\np)_{r_1r_2}
  \label{line3}\\
   & &
\kern -2cm =
\frac{(2\pi)^3}{V^2}
\delta(\nh+\nq-\np)
\int \dif p_1 \dif p_2 \sum_{r_1r_2}
  j_a^\mu(\np s_p,\nk s_k; \np_1 r_1, \np_2 r_2)
  \delta(\np_1+\np_2-\nh-\nk)
  \phi_{\nh\nk}^{s_hs_k}(\np)_{r_1r_2}
  \nonumber\\
   & &
\kern -2cm =
  \frac{(2\pi)^3}{V^2}
\delta(\nh+\nq-\np)
\int \dif p_1 \sum_{r_1r_2}
  j_a^\mu(\np s_p,\nk s_k; \np_1 r_1, \np_2 r_2)
  \phi_{\nh\nk}^{s_hs_k}(\np)_{r_1r_2}
  \end{eqnarray}
where $\np=(\np_1-\np_2)/2$ is the relative momentum of $\np_1,\np_2$,
and $\np_2=\nh+\nk-\np_1$ by momentum conservation.  In Eq. (\ref{line2})
 we have interchanged the brackets between $|p_1p_2\rangle$ and
$\Phi_{hk}$. In Eq. (\ref{line3})  we applied the two--body operator
matrix element, Eq.~(\ref{TBmatrix}), and expressed the correlated
wave function as the product of the total momentum plane wave and the
relative function $\phi_{hk}$. As a consequence, the total momentum
$\nh+\nk$ is preserved because
\begin{equation}
  \langle \np_1 r_1, \np_2 r_2|\Phi_{\nh\nk}^{s_hs_k}\rangle=
  \delta(\np_1+\np_2-\nh-\nk)
  \phi_{\nh\nk}^{s_hs_k}(\np)_{r_1r_2}.
\end{equation}
It is worth pointing out that the relative wave function
$\phi_{\nh\nk}^{s_hs_k}$ is actually a bi--spinor, and therefore its
spinorial components are written as subscripts,
$\phi_{\nh\nk}^{s_hs_k}(p)_{r_1r_2}$. The superscripts $s_h$ and $s_k$
refer to the spins of the uncorrelated pair before the interaction.

\end{widetext}


\end{document}